\documentclass[aps,prd,12pt,a4paper,groupedaddress,preprintnumbers,showpacs,floatfix,nofootinbib]{revtex4}
\usepackage[english]{babel}
\usepackage{amsmath}
\usepackage{graphicx}
\usepackage{color}
\usepackage{amssymb}
\usepackage{hyperref}
\usepackage{dcolumn}% Align table columns on decimal point
\usepackage{bm}% bold math
\newcommand{\buno}{\mathbf{1}}
\newcommand{\bdos}{\mathbf{2}}

\begin{document}

\preprint{DCP-11-01}

%opening
\title{Model of flavor with quaternion symmetry}

\author{
 Alfredo Aranda,$^{1,2,3}$\footnote{Electronic address:fefo@ucol.mx}
 Cesar Bonilla,$^{1}$\footnote{Electronic address:rasec.cmbd@gmail.com}
 Raymundo Ramos,$^{1}$\footnote{Electronic address:raalraan@gmail.com}
 and Alma D. Rojas$^{1}$\footnote{Electronic address:alma.drp@gmail.com}}

\affiliation{$^1$Facultad de Ciencias, CUICBAS,
Universidad de Colima, Colima, M\'exico \\
$^2$Dual C-P Institute of High Energy Physics, M\'exico \\
$^3$Abdus Salam ICTP, Trieste, Italy}

\date{\today}

\begin{abstract}
We present a renormalizable fermion mass model based on the symmetry $Q_4$ that accommodates all fermion masses and mixing angles in both the quark and lepton sectors. It requires the presence of only four SU(2) doublet scalar fields transforming non trivially under the flavor symmetry and the assumption of an alignment between first and second generation Yukawa couplings. No right-handed neutrinos are present in the model and neutrino masses are generated radiatively through the introduction of two additional SU(2) singlet fields charged under both hypercharge and lepton number.
\end{abstract}


\maketitle

%%%%%%%%%%%%%%%%%%%%%%%%%%%%%%%%%%%%%%%%%%%%%%%%%%%%%%%%%%%%%%%%%%%%%%%%%%%%%%%%%%%%%%%%
%%%%%%%%%%%%%%%%%%%%%%%%%%%%%%%%%%%%%%%%%%%%%%%%%%%%%%%%%%%%%%%%%%%%%%%%%%%%%%%%%%%%%%%%
%%%%%%%%%%%%%%%%%%%%%%%%%%%%%%%%%%%%%%%%%%%%%%%%%%%%%%%%%%%%%%%%%%%%%%%%%%%%%%%%%%%%%%%%
\section{Introduction}
\label{introduction}
Discrete symmetries have been used extensively in models of fermion masses for several years now (for an extensive list of references please see~\cite{Altarelli:2010gt,Ishimori:2010au,Altarelli:2010fk}). The finiteness in their number of representations lets one imagine the possibility of a predictive scenario. Their actual implementation into a realistic model, however, usually comes with a plethora of assumptions and additions to the Standard Model (SM) that, depending on the specific setup and ambition, may or not be experimentally testable. 

The recent results on neutrino mixing angles and mass squared differences have given more impetus to the flavor model builders, particularly the observation that the neutrino mixing angles closely match the so-called tribimaximal mixing~\cite{tribimaximal}. This observation alone has led to a close analysis on the symmetry properties needed (inherent) in the lepton sector~\cite{lametc}. Among the most popular - and prolific - groups explored in this regard is $A_4$, the group of even permutations on four elements (same as $T$, the group of orientation-preserving symmetries of a tetrahedron)~\cite{A4}.

It is not a settled matter whether or not quarks and leptons are both touched by the same flavor symmetry. On the one hand the {\it unexpected} maximality in the neutrino mixing 
sector~\footnote{Most work done before the maximal mixing was determined focused on the small mixing angle solutions. Some exceptions can be found 
in~\cite{Cabibbo:1977nk,Wolfenstein:1978uw,Ma:2009wt}.}, compared to the hierarchical and small mixing observed in the quark sector, could be an indication that they should be treated independently. On the other hand models that incorporate a single symmetry in both sectors do exist and thus, from a model building perspective, it is certainly possible to have both sectors connected through a single flavor symmetry. This last possibility can also be motivated in grand unified scenarios.

Models in this category contain a large number of additional fields to those of the SM (or other frameworks such as the Minimal Supersymmetric Standard Model (MSSM), Grand Unified Theories (GUTs), etc.). Most of these new additional fields are scalar fields needed to break the flavor symmetry and/or to generate hierarchies through ratios of their vacuum expectation values (vevs) to high energy flavor scales. In most cases these so-called flavon fields are taken to be heavy and do not lead to detectable phenomenology (for a study of possible flavon effects at the LHC see~\cite{liliana}). A possible alternative to this situation is provided by renormalizable flavor models in which the scalars responsible for electroweak symmetry breaking (EWSB) are also charged under the flavor symmetry~\cite{renormalizable,Morisi:2010rk}. In such a scenario the scalar fields may have significant phenomenology at accessible energy scales.

Neutrino mass generation plays an important role on both approaches. The smallness of neutrino masses has to be attributed to some additional mechanism that must be incorporated into the models, the seesaw being the most popular and perhaps successful~\cite{seesaw}. The end result is the need to add more scalars and/or energy scales. In some renormalizable models non-renormalizable operators are introduced to generate neutrino masses and this requires also the introduction of some high energy scale (without the introduction of right-handed neutrinos). As a side note we mention that for models with right-handed neutrinos, there has been a recent interest in the possibility of lowering down the scale associated to neutrino mass generation close to the electroweak (EW) scale and thus, perhaps, make it accessible to experiments at the Large Hadron Collider (LHC). See~\cite{ewneutrinos} for an incomplete list of examples.

The two general approaches described above are then useful and have led to interesting possibilities. If one is interested in explaining the observed hierarchies in the masses without assuming a hierarchical structure for the Yukawa couplings, then the first (flavon based) approach seems appropriate at the expense of introducing high energy - unobservable - scales. If one is instead interested in the possibility of accessing the phenomenology associated to a possible flavor model, then the second approach may seem more appropriate - at the expense of assuming hierarchical couplings. Nothing is for free.

As mentioned above, one of the attractive features of renormalizable models is that the SU(2) doublet scalar fields transform non-trivially (at least some of them) under the flavor symmetry and this can, in principle, be reflected phenomenologically. Most models however require a large number of SU(2) doublets (and sometimes triplets) in order to obtain realistic fermion mass matrices and mixing angles. Most models have in their construction the strong requirement for the symmetry to determine the tribimaximality in the lepton sector. The quark sector is then accommodated through the incorporation of more scalars and/or additional Abelian symmetries. An interesting question is to determine if it is possible to create a renormalizable model with a few (compared to $\geq 7$ for models in the literature) SU(2) doublet scalar fields that would in principle lead to interesting - more tractable - EW scale phenomenology. 

In this work we address this question and find that, under certain conditions, it is possible to create models with a minimum of four SU(2) doublet scalar fields. The starting point for our approach relies on the study of the Fritzsch - like textures~\cite{Fritzsch:1977vd} in the quark sector in order to determine which groups can be used to reproduce them with the minimum number of SU(2) doublets (we only consider non-supersymmetric models). Once this is determined, the charged lepton sector can be obtained automatically in analogy with the down-type quarks - note however that this determines the representation of left-handed neutrinos under the flavor symmetry and so it must be checked whether or not that same representation leads to acceptable results for neutrino mass differences and mixing angles (in general it does not!). As for the neutrino sector, the models do not include right-handed neutrinos and masses are generated radiatively~\cite{Zee}. In order to accomplish this at least two additional SU(2) singlet scalar fields are needed with non-zero hypercharge, charged under Lepton number, and with non-trivial representations under the flavor symmetry. The smallest group we find that can be used in this scenario is the quaternion group $Q_4$ and a model based on it is presented in detail. The model successfully accommodates all data on both quark and lepton sectors only for an inverted hierarchy in the neutrino sector and without exact tribimaximality ($Q_4$ has been used before as a flavor symmetry in different scenarios, see for example~\cite{q4models}).   

In Section~\ref{model} we present the general description of the model based on $Q_4$ including the results and discussion of the numerical analysis. The scalar potential and vacuum alignment for the model is discussed in Section~\ref{potential}. The phenomenological study associated to the scalar sector is under investigation and will be presented in another publication. We then present our conclusions and final remarks. We have included three appendices where we give some details on the group $Q_4$, the analysis of the Yukawa mass matrices in the quark sector, and finally the radiative generation of neutrino masses.

%%%%%%%%%%%%%%%%%%%%%%%%%%%%%%%%%%%%%%%%%%%%%%%%%%%%%%%%%%%%%%%%%%%%%%%%%%%%%%%%%%%%%%%%
%%%%%%%%%%%%%%%%%%%%%%%%%%%%%%%%%%%%%%%%%%%%%%%%%%%%%%%%%%%%%%%%%%%%%%%%%%%%%%%%%%%%%%%%
%%%%%%%%%%%%%%%%%%%%%%%%%%%%%%%%%%%%%%%%%%%%%%%%%%%%%%%%%%%%%%%%%%%%%%%%%%%%%%%%%%%%%%%%

\section{The model}
\label{model}

Consider the SM gauge and fermion content plus four additional SU(2) scalar doublets ($H_i$, $i=1,2,3,4$) and two SU(2) singlet scalar fields ($\eta_1$ and $\eta_2$) with hypercharge $Y=-1$ and Lepton number $L=2$ (note that no right-handed neutrinos are present). Now assume there is an additional flavor symmetry $Q_4$ under which the fields above transform in the following way:
\begin{eqnarray}\label{transformations}
\overline{Q} &\sim& \buno^{++} \oplus \buno^{+-} \oplus \buno^{-+} \equiv \{ \overline{Q}_1 \oplus \overline{Q}_2 \oplus \overline{Q}_3 \} \nonumber  \\
d_R &\sim& \bdos \oplus \buno^{+-} \equiv  \{ (d_{R1} \ \ d_{R2}) \oplus d_{R3} \} \nonumber\\
u_R &\sim& \bdos \oplus \buno^{+-} \equiv  \{ (u_{R2} \ \ u_{R1}) \oplus u_{R3} \} \nonumber\\
\overline{L} &\sim& \bdos \oplus \buno^{+-} \equiv  \{ (\overline{L}_1 \ \ \overline{L}_2 ) \oplus 
\overline{L}_3\} \\
e_R &\sim& \buno^{++} \oplus \buno^{+-} \oplus \buno^{-+} \equiv  \{ e_{R1} \oplus e_{R2} \oplus e_{R3} \} \nonumber\\
H &\sim& \bdos \oplus \buno^{++} \oplus \buno^{--} \equiv \{ H_D \equiv (H_1 \ \ H_2) \oplus H_3 \oplus H_4 \} \nonumber\\
\eta &\sim& \bdos \equiv \{ \eta_D \equiv (\eta_1 \ \ \eta_2) \} \nonumber\ ,
\end{eqnarray}
where $Q$ and $L$ denote the SU(2) doublets for left-handed quarks and leptons respectively and fields with subscript $R$ denote SU(2) singlet right-handed fermion fields. Note in particular that the ordering of first and second generation fields in the doublet of $Q_4$ for the right-handed up-type quarks is reversed compared to the down-type quarks. This is necessary in order to obtain the same texture in both sectors (see Appendix~\ref{textures} for a possible alternative). This is interesting since one naively could expect that the ordering of families should have no effect, i.e. it would amount to a basis rotation. Nevertheless, the non-trivial transformation under the flavor symmetry does produce an effect~\cite{fefo-tprime}. Another thing to note is the difference in representations between $Q$ and $L$. We alluded to this in the Introduction and as it turns out, letting $L \sim \buno \oplus \buno \oplus \buno$ does not accommodate acceptable results in the neutrino sector. This is an interesting result that shows that the symmetry does play a role in the determination of the mixing angles and mass differences in the neutrino sector as well.

We now present the consequences of these charge assignments for the quarks and lepton mass matrices.

%%%%%%%%%%%%%%%%%%%%%%%%%%%%%%%%%%%%%%%%%%%%%%%%%%%%%%%%%%%%%%%%%%%%%%%%%%%%%%%%%%%%%%%%
%%%%%%%%%%%%%%%%%%%%%%%%%%%%%%%%%%%%%%%%%%%%%%%%%%%%%%%%%%%%%%%%%%%%%%%%%%%%%%%%%%%%%%%%
%%%%%%%%%%%%%%%%%%%%%%%%%%%%%%%%%%%%%%%%%%%%%%%%%%%%%%%%%%%%%%%%%%%%%%%%%%%%%%%%%%%%%%%%

\subsection{Quark sector}
\label{quarksector}
The down-type quark mass matrices are obtained from the following gauge and flavor invariant terms (see Appendix~\ref{q4details}):
\begin{equation}
Y^{d}_{0}\overline{Q}_{1}d_{DR}H_{D}= \buno^{++} \otimes \underbrace{\bdos\otimes \bdos}_{\supset\buno^{++}}\supset\buno^{++}=Y^{d}_{0}
\overline{Q_{1}}d_{2R}H_{1}-Y^{d}_{0}\overline{Q_{1}}d_{1R}H_{2},
\end{equation}

\begin{equation}
Y^{d}_{1}\overline{Q}_{2}d_{DR}H_{D}= \buno^{+-} \otimes \underbrace{\bdos\otimes \bdos}_{\supset\buno^{+-}} \supset \buno^{++}=Y^{d}_{1}
\overline{Q_{2}}d_{1R}H_{1}-Y^{d}_{1}\overline{Q_{2}}d_{2R}H_{2},
\end{equation}

\begin{equation}
Y^{d}_{2}\overline{Q}_{3}d_{DR}H_{D}= \buno^{-+} \otimes \underbrace{\bdos\otimes \bdos}_{\supset\buno^{-+}} \supset \buno^{++}=Y^{d}_{2}
\overline{Q_{3}}d_{1R}H_{2}+Y^{d}_{2}\overline{Q_{3}}d_{2R}H_{1},
\end{equation}

\begin{equation}
Y^{d}_{3}\overline{Q}_{2}d_{3R}H_{3}\sim \buno^{+-}\otimes \buno^{+-} \otimes \buno^{++}\supset\buno^{++},
\end{equation}

\begin{equation}
Y^{d}_{4}\overline{Q}_{3}d_{3R}H_{4}\sim \buno^{-+}\otimes \buno^{+-} \otimes \buno^{--}\supset\buno^{++},
\end{equation}
where the $Q_4$ products are shown explicitly and where the $Y_a^d$ represent numerical unknown coefficients to be determined later. We have omitted their hermitian conjugates for simplicity.   

After electroweak symmetry breaking (EWSB), which we assume is triggered by the CP-even vacuum expectation values (vevs) of the SU(2) doublets $H_i$, the following mass matrix is obtained:
\begin{eqnarray}\label{md}
M_{d}= \left( \begin{array}{ccc}
-Y^{d}_{0} v_{2}& Y^{d}_{0} v_{1} & 0 \\
Y^{d}_{1} v_{1}& -Y^{d}_{1} v_{2} & Y^{d}_{3} v_{3}\\
Y^{d}_{2} v_{2}& Y^{d}_{2} v_{1} & Y^{d}_{4} v_{4} 
\end{array} \right) \ ,
\end{eqnarray}
where the $v_{i}$s denote the vevs $\langle H_{i}\rangle=v_{i}$.

For the up-type quark sector we obtain
\begin{equation}
Y^{u}_{0}\overline{Q}_{1}u_{DR}\tilde{H}_{D}= \buno^{++} \otimes \underbrace{\bdos\otimes \bdos}_{\supset\buno^{++}} \supset \buno^{++}=Y^{u}_{0}\overline{Q}_{1}u_{1R}\tilde{H}_{2}+Y^{u}_{0}\overline{Q}_{1}u_{2R}\tilde{H}_{1},
\end{equation}

\begin{equation}
Y^{u}_{1}\overline{Q}_{2}u_{DR}\tilde{H}_{D}= \buno^{+-} \otimes \underbrace{\bdos\otimes \bdos}_{\supset\buno^{+-}}\supset \buno^{++}=Y^{u}_{1}
\overline{Q_{2}}u_{2R}\tilde{H}_{2}+Y^{u}_{1}\overline{Q_{2}}u_{1R}\tilde{H}_{1},
\end{equation}

\begin{equation}
Y^{u}_{2}\overline{Q}_{3}u_{DR}\tilde{H}_{D}= \buno^{-+} \otimes \underbrace{\bdos\otimes \bdos}_{\supset\buno^{-+}}\supset \buno^{++}=Y^{d}_{2}
\overline{Q_{3}}u_{2R}\tilde{H}_{1}-Y^{d}_{2}\overline{Q_{3}}u_{1R}\tilde{H}_{2},
\end{equation}

\begin{equation}
Y^{u}_{3}\overline{Q}_{2}u_{3R}\tilde{H}_{3}\sim \buno^{+-}\otimes \buno^{+-} \otimes \buno^{++}\supset \buno^{++},
\end{equation}

\begin{equation}
Y^{u}_{4}\overline{Q}_{3}u_{3R}\tilde{H}_{4}\sim \buno^{-+}\otimes \buno^{+-} \otimes \buno^{--}\supset \buno^{++},
\end{equation}
where $\tilde{H} \equiv i\sigma_2H^*$. After EWSB these expressions lead to
\begin{eqnarray} \label{mu}
M_{u}= \left( \begin{array}{ccc}
Y^{u}_{0} v_{2}& Y^{u}_{0} v_{1} & 0 \\
Y^{u}_{1} v_{1}& Y^{u}_{1} v_{2} & Y^{u}_{3} v_{3}\\
-Y^{u}_{2} v_{2}& Y^{u}_{2} v_{1} & Y^{u}_{4} v_{4} 
\end{array} \right) \ .
\end{eqnarray}

In order to obtain a Fritzsch-like pattern for these matrices the following assumptions are made: $Y_0^{d} = Y_1^{d}$, $Y_0^{u} = Y_1^{u}$, and $v_2 = 0$. We were not able to obtain the condition on the unknown coefficients from the flavor symmetry without enlarging the model by using larger groups and more scalars, and thus it is our strongest assumption. The vacuum alignment condition is analyzed in section~\ref{potential} where it is shown that it is consistent with vacuum stability.

Under these assumptions the mass matrices acquire the following textures:
\begin{eqnarray}\label{QMM}
M_{u,d} = \left( \begin{array}{ccc}
0 & A_{u,d} & 0 \\
A_{u,d} & 0 & B_{u,d} \\
0 & D_{u,d} & C_{u,d}
\end{array} \right),
\end{eqnarray}
where we have parametrized the products of the unknown coefficients $Y$ with the vevs in terms of the new coefficients $A$, $B$, $C$, and $D$.

Following the analysis presented in~\cite{Morisi:2010rk,nuestro}
and taking $C_{u,d}=y^{2}_{u,d} m_{t,b}$, we rewrite the mass matrices above in terms of the quark masses and free parameters $y_{u,d}$~\cite{Babu:2004tn},
\begin{eqnarray}\label{mumdfinal}
\hat{M}_{u,d} =m_{t,b} \left( \begin{array}{ccc}
0 & q_{u,d}/y_{u,d} & 0 \\
q_{u,d}/y_{u,d} & 0 & b_{u,d} \\
0 & d_{u,d} & y^{2}_{u,d} 
\end{array} \right) \ ,
\end{eqnarray}
where
\begin{eqnarray}
q^{2}_{u,d} = \frac{m_{u,d}m_{c,s}}{m_{t,b}^2},
\end{eqnarray}
\begin{eqnarray}
p_{u,d} = \frac{m_{u,d}^2 + m_{c,s}^2}{m_{t,b}^{2}},
\end{eqnarray}
\begin{eqnarray}
d_{u,d} =  \sqrt{\frac{p_{u,d} + 1 - y_{u,d}^{4} + R_{u,d}}{2} - \left(\frac{q_{u,d}}{y_{u,d}}\right)^2},
\end{eqnarray}
\begin{eqnarray}
b_{u,d} = \sqrt{\frac{p_{u,d} + 1 - y_{u,d}^{4} - R_{u,d}}{2} - \left(\frac{q_{u,d}}{y_{u,d}}\right)^2},
\end{eqnarray}
\begin{eqnarray}
R_{u,d} = ((1 + p_{u,d} - y_{u,d}^{4})^{2} - 4(p_{u,d} + q_{u,d}^{4}) + 8q_{u,d}^{2}y_{u,d}^{2})^{1/2},
\end{eqnarray}
and where $\hat{M}_{u,d}$ are matrices with real entries obtained from the phase factorization
of $M_{u,d}$~\cite{Babu:2004tn} through
\begin{eqnarray}\label{quarkphases}
M_{u,d}=P_{u,d}^{\ast}\hat{M}P_{u,d}
\end{eqnarray}
with $P_{u,d}$ diagonal phase matrices such that
$P=P_{u}P_{d}^{\ast}=\operatorname{diag}(1,e^{i\beta_{ud}},e^{i\alpha_{ud}})$ with $\beta_{ud} \equiv \beta_u-\beta_d$ and  $\alpha_{ud} \equiv \alpha_u-\alpha_d$.

The free parameters are then $y_{u,d}$, $\alpha_{ud}$, and $\beta_{ud}$, and the CKM matrix is given by
\begin{eqnarray}
V_{CKM}=\mathcal{O}_{u}P\mathcal{O}^{T}_{d},
\end{eqnarray}
where the $\mathcal{O}_{u,d}$ matrices diagonalize $\hat{M}^{2}_{u,d}$ via,
\begin{eqnarray}
\mathcal{O}_{u,d}\hat{M}_{u,d}\hat{M}^{T}_{u,d}\mathcal{O}^{T}_{u,d}=\operatorname{diag}(m^{2}_{u,d},m^{2}_{c,s},m^{2}_{t,b}).
\end{eqnarray}

Using the values $y_u=0.9964$, $y_d=0.9623$, $\alpha_{ud}=1.9560$, and $\beta_{ud}=1.4675$ we obtain
\begin{eqnarray} \nonumber
V^{th}_{CKM} = \mathcal{O}_uP \mathcal{O}_d^{T} =
\left(
\begin{array}{ccc}
 0.97434 +i0.00976  & -0.22086+i0.0422  & 0.0035 -i0.00098 \\
 -0.0197+i0.2239 & 0.10837 +i0.9675 & 0.03395 -i0.02179 \\
 0.00676 +i0.00505 & 0.0258 +i0.03006 & -0.373764+i0.92664
\end{array}
\right),
\end{eqnarray}
and so
\begin{eqnarray}
|V^{th}_{CKM}|=\left( \begin{array}{ccc}
0.974386 & 0.224853 & 0.00363\\
0.224723  & 0.973587 & 0.0403354 \\
0.00844 & 0.0396092  & 0.99918
\end{array} \right) \ ,
\end{eqnarray}
and~\footnote{We compute $\delta^{th}_{CKM}$ using the expressions in~\cite{Antusch:2009hq}} $\delta^{th}_{CKM}=1.19528$ in agreement with the experimental data~\cite{PDG2010}
\begin{eqnarray}
|V_{CKM}| =\left( \begin{array}{ccc}
0.97428\pm0.00015 & 0.2253 \pm 0.0007   & 0.00347 ^{+0.00016} _{-0.00012} \\
0.2252 \pm {0.0007}    & 0.97345^{+0.00015}_{-0.00016}  & 0.0410 ^{+0.0011}_{-0.0007} \\
0.00862^{+0.00026}_{-0.00020} & 0.0403 ^{+0.0010}_{-0.0007}  & 0.999152^{+0.000030}_{-0.000045}
\end{array} \right) \ ,
\end{eqnarray}
and $\delta_{CKM}= 1.20146^{+0.04758}_{-0.06963}$.

%%%%%%%%%%%%%%%%%%%%%%%%%%%%%%%%%%%%%%%%%%%%%%%%%%%%%%%%%%%%%%%%%%%%%%%%%%%%%%%%%%%%%%%%
%%%%%%%%%%%%%%%%%%%%%%%%%%%%%%%%%%%%%%%%%%%%%%%%%%%%%%%%%%%%%%%%%%%%%%%%%%%%%%%%%%%%%%%%
%%%%%%%%%%%%%%%%%%%%%%%%%%%%%%%%%%%%%%%%%%%%%%%%%%%%%%%%%%%%%%%%%%%%%%%%%%%%%%%%%%%%%%%%

\subsection{Lepton sector}
\label{leptonsector}

In this case the allowed Yukawa terms are

\begin{equation}
Y^{\ell}_{0}\overline{L}_{D}e_{R_{1}}H_{D}= \bdos \otimes \buno^{++} \otimes \bdos \supset \buno^{++}=Y^{\ell}_{0}
\overline{L}_{1}e_{1R}H_{2}-Y^{\ell}_{0}\overline{L}_{2}e_{1R}H_{1},
\end{equation}

\begin{equation}
Y^{\ell}_{1}\overline{L}_{D}e_{R_{2}}H_{D}= \bdos \otimes \buno^{+-}\otimes \bdos \supset \buno^{++}=Y^{\ell}_{1}
\overline{L}_{1}e_{2R}H_{1}-Y^{\ell}_{1}\overline{L_{2}}e_{2R}H_{2},
\end{equation}

\begin{equation}
Y^{\ell}_{2}\overline{L}_{D}e_{R_{3}}H_{D}= \bdos \otimes\buno^{-+}\otimes \bdos \supset \buno^{++}=-Y^{\ell}_{2}
\overline{L}_{1}e_{3R}H_{2}-Y^{\ell}_{2}\overline{L_{2}}e_{3R}H_{1},
\end{equation}

\begin{equation}
Y^{\ell}_{3}\overline{L}_{3}e_{2R}H_{3}\sim \buno^{+-}\otimes \buno^{+-} \otimes \buno^{++}\supset\buno^{++},
\end{equation}

\begin{equation}
Y^{\ell}_{4}\overline{L}_{3}e_{3R}H_{4}\sim \buno^{+-}\otimes \buno^{-+} \otimes \buno^{--}\supset\buno^{++},
\end{equation}
which written in matrix form, after EWSB, gives
\begin{eqnarray}
M_{d}= \left( \begin{array}{ccc}
 Y^{d}_{0} v_{2}  &  Y^{d}_{1} v_{1}   & -Y^{d}_{2} v_{2} \\
-Y^{d}_{0} v_{1}  & -Y^{d}_{1} v_{2}   & -Y^{d}_{2} v_{1}\\
 0                &  Y^{d}_{3} v_{3}   &  Y^{d}_{4} v_{4} 
\end{array} \right) \ .
\end{eqnarray}
The analysis made for the quark sector extends directly to the charged leptons. The mass matrix can then be written as before (see Eq.~(\ref{mumdfinal})):
\begin{eqnarray}\label{mlfinal}
\hat{M}_{l} =m_{\tau} \left( \begin{array}{ccc}
0 & a_{l} & 0 \\
-a_{l}& 0 & b_{l} \\
0 & d_{l} & y^{2}_{l}
\end{array} \right)  \ ,
\end{eqnarray}
where again we have made the assumptions that $Y_0^d = Y_1^d$ and $v_2=0$.

Calling $U_l$ the matrix that diagonalizes $\hat{M}_l^2$, and using the values for the charged lepton masses taken from~\cite{PDG2010}, we obtain  
\begin{eqnarray}\label{Ul th}
   U_l= \left(%
\begin{array}{ccc}
 0.997042  & 0.0624654 & -0.0447713 \\
  0.0768522 &-0.813271  & 0.576787 \\
  -0.000382008 & -0.578522  &  -0.815667\\
\end{array}%
\right),
\end{eqnarray}
with $y_l = 0.9$ and $\alpha_l=\beta_l=0$.

For neutrinos the situation is different. We are assuming that neutrinos are Majorana type and that their masses can be induced by radiative corrections, thus making them light naturally~\cite{Babu:1988qv}. Following the description in Appendix~\ref{neutrinos} we write the symmetry allowed interactions of the fields $\eta_1$ and $\eta_2$ with leptons and with the SU(2) scalars. For leptons the interaction terms are given by
\begin{equation}
\mathcal{L}_{LLh}=\kappa \epsilon_{ij} \overline{L_{iD}^c}L_{j\tau} \eta_{D}^* + h.c.
\label{llh}
\end{equation}
where $i,j=1,2$ are $SU(2)$ indices, $\epsilon_{ij}=-\epsilon_{ji}=1$, and $\kappa$ denotes the antisymmetric matrix (in family space) 
\begin{equation}
\kappa=\left(\begin{array}{ccc}
0&0&\kappa_{D}\\
0&0&\kappa_{D}\\
-\kappa_{D}&-\kappa_{D}&0
\end{array}\right)\label{kappa}
\end{equation}
with $\kappa_D$ a free parameter that characterizes the {\it size} of the interaction. 

The gauge invariant - Lepton number violating - interaction terms with the SU(2) scalars are given by (see Eq.~(\ref{Lcubic}))
\begin{eqnarray}\label{vHHeta}
V_{HH\eta}&=&\lambda_1 \epsilon_{ij} H_{Di} H_{3j} \eta_{D}+\lambda_2 \epsilon_{ij} H_{Di} H_{4j} \eta_{D} + h.c.\nonumber \\
&=&\lambda_1 (\epsilon_{ij} H_{1i} H_{3j}\eta_{2}-\epsilon_{ij} H_{2i} H_{3j}\eta_{1})+\lambda_2 (\epsilon_{ij} H_{1i} H_{4j}\eta_{1}+\epsilon_{ij} H_{2i} H_{4j}\eta_{2}) + h.c.  \ .
\end{eqnarray}

We compute the neutrino mass matrix elements by evaluating diagrams like the one in Figure~\ref{fig:loop}. Consider the fermion line in such diagram. Its contribution is given by 
\begin{equation}
U_L Y_l U_R^\dag U_R M_l^\dag U_L^\dag U_L \kappa U_L^\dag =
U_L Y_l  M_l^\dag \kappa U_L^\dag \sim U_L M_{\nu} U_L^{\dagger} \equiv M_{\nu}^{\prime}\ ,
\label{uymku}
\end{equation}
where
\begin{equation}
U_L M_lU_R^\dag =\text{diag}(m_e,m_\mu,m_\tau)  \ ,
\label{diag}
\end{equation}
$Y_l$ is the lepton Yukawa matrix
\begin{equation}
Y_l=\left(\begin{array}{ccc}
-Y&Y&Y_2\\
Y&-Y&Y_2\\
0&Y_3&Y_{4}
\end{array}\right)\label{yl} \ ,
\end{equation}
and where the matrices $M_{\nu}^{\prime}$ and $M_{\nu}=Y_lM_l^{\dagger}\kappa$ correspond to the neutrino mass matrices in the charged lepton mass and weak bases respectively (up to factors from the scalar loops). The neutrino mixing is then obtained by diagonalizing $M_{\nu}^{\prime}$ using~\cite{BenTov:2011tj}
\begin{eqnarray}
M_{\nu}^{\prime} = V^*M_{\nu}^DV^{\dagger} \ ,
\end{eqnarray}
where $V={\cal K} V_{PMNS} {\cal M}$ with $V_{PMNS}\equiv U_L^{\dagger}U_{\nu}$, ${\cal K}\equiv 
diag(e^{i\kappa_1},e^{i\kappa_2},e^{i\kappa_3})$, ${\cal M} \equiv diag(e^{i\sigma},e^{i\rho},1)$, and $M_{\nu}^D$ representing the diagonal neutrino mass matrix with eigenvalues $m_i \geq 0$ (corresponding to the physical neutrino masses). The phases $\kappa_i$ are unphysical and in our analysis we set them to zero. The phases $\sigma$ and $\rho$ are Majorana phases that are determined from the diagonalization. We use the standard parametrization for $V_{PMNS}$ namely:
\begin{eqnarray}
V_{PMNS} = 
\left(
\begin{array}{ccc}
-c_{13}c_{12} & c_{13}s_{12} & s_{13} e^{-i\delta_{CP}} \\
c_{23}s_{12}+s_{23}s_{13}c_{12}e^{i\delta_{CP}} & c_{23}c_{12} - s_{23}s_{13}s_{12}e^{i\delta_{CP}}  & s_{23}c_{13} \\ 
s_{23}s_{12} -c_{23}s_{13}c_{12}e^{i\delta_{CP}} & s_{23}c_{12}+c_{23}s_{13}s_{12}e^{i\delta_{CP}} & -c_{23}c_{13}
\end{array}
\right) \ ,
\end{eqnarray}
where $c_{ij} \equiv \cos\theta_{ij}$, $s_{ij}\equiv \sin\theta_{ij}$, and $\delta_{CP}$ is the CP-violating phase (we assume that CP is conserved in this sector and thus work with $\delta_{CP}=0$).

Using Eq.~(\ref{uymku}) we obtain the matrix elements for $M_{\nu}$ (including now the scalar loop factors)
\begin{eqnarray}\label{neutrinomasses}
m_{\nu_e\nu_e}&=& -m_{\nu_\mu\nu_\mu} = (Y\kappa_D \lambda_2 m_{\tau\mu}v_{4}-Y_2\kappa_{D} \lambda_1 m_{\tau\tau} v_{3})F(m_H^2,m_{\eta}^2)\\
%--------------------------------------
%m_{\nu_\mu\nu_\mu}&=&(-Y\kappa_D \lambda_2 m_{\tau\mu}v_{4}+Y_2\kappa_{D} \lambda_1 m_{\tau\tau} v_{3})F(m_H^2,m_{\eta}^2)\\
%--------------------------------------
m_{\nu_e\nu_\mu}&=&m_{\nu_\mu\nu_e}=(2 \kappa_D m_{\tau\mu}\lambda_1 Y v_{3}+2 \kappa_{D} \lambda_2 m_{\tau\tau} Y_2 v_{4}) F(m_H^2,m_{\eta}^2)\\
%--------------------------------------
m_{\nu_e\nu_\tau}&=&m_{\nu_\tau\nu_e}=(-\kappa_{D} \lambda_2 m_{\tau\tau} Y_{4} v_{1}-\kappa_{D} \lambda_2 m_{\mu\tau} Y_{2} v_{4}) F(m_H^2,m_{\eta}^2)\\
%--------------------------------------
m_{\nu_\mu\nu_\tau}&=&m_{\nu_\tau\nu_\mu}=-(\kappa_{D} \lambda_1 m_{\mu\tau} Y_{2} v_{3}+\kappa_{D} \lambda_1 m_{\tau\mu} Y_{3} v_{1}+2\kappa_{D} \lambda_1 m_{e\mu} Y v_{3}) F(m_H^2,m_{\eta}^2).
\end{eqnarray}

Assuming that $\lambda_1 \sim m_{H^+}\sim 500$~GeV, and noting that
$Y\langle H\rangle$ must be at the same scale of $m_l$, then if $\kappa_D\sim$~O(1) (O($10^{-3}$)) then $m_{\eta}\sim 4\times 10^5$~GeV ($9 \times 10^3$~GeV)
leads to matrix elements of O(eV).

The Majorana neutrino mass matrix then has the texture:
\begin{equation}
\label{mnu}
M_\nu= \left(\begin{array}{ccc}
a&c&d\\
c&-a&e\\
d&e&0\\
\end{array}\right) \ ,
\end{equation}
where all entries are O(eV).

In order to perform the numerical analysis we use the following experimental results~\cite{PDG2010}:
\begin{eqnarray}\label{lepton-angles}
\sin^2(2\theta_{12}) & = & 0.087 \pm 0.03 \\
\sin^2(2\theta_{23}) & > & 0.92 \\
\sin^2(2\theta_{13}) & < & 0.15 
\end{eqnarray}
and
\begin{eqnarray}\label{lepton-masses}
\Delta m_{21}^2 & = & 7.59_{-0.21}^{+0.19}\times 10^{-5} \ \rm{eV}^2 \\
\Delta m_{32}^2 & = & 2.43 \pm 0.13\times 10^{-3} \ \rm{eV}^2 \ .
\end{eqnarray}
Since the absolute mass scale in the neutrino sector is not known, we use the following ratio
\begin{eqnarray}\label{mass-ratio}
0.0338 < \left| \frac{\Delta m_{21}^2}{\Delta m_{32}^2} \right| < 0.0288 \ .
\end{eqnarray}

In order to determine whether the mass matrices in this model can reproduce these results, we performed a scan of the complete range in all three angles. Then for each case where a solution consistent with all three angles was found, we computed the ratio in Eq.~(\ref{mass-ratio}) and selected those solutions that fell within its allowed range. We found that solutions exist with the following properties:

\begin{enumerate}
\item Solutions exist only for an inverted hierarchy ($m_3\ll m_1\approx m_2$),
\item The mixing angles are bounded by
\begin{eqnarray}
0.84&<&\sin^2(2\theta_{12})<0.9 \ , \\
0.96&<&\sin^2(2\theta_{23})<1 \ , \\
0.012&<&\sin^2(2\theta_{13})<0.15 \ .
\end{eqnarray}
\item Note that while for $\theta_{12}$ there are consistent solutions for all the experimental range, the angles $\theta_{13}$ and $\theta_{23}$ have an inferior bound higher than the experimental. It can be seen that the model always deviates from exact tribimaximal mixing.
\end{enumerate}

\begin{figure}[ht]
\includegraphics[width=8cm]{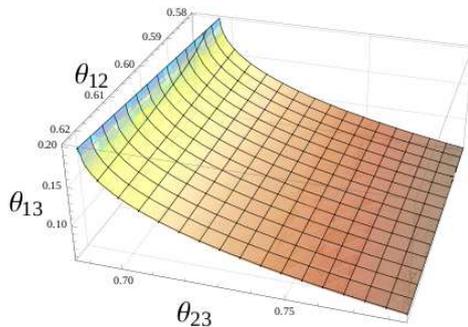}
\caption{Angles that reproduce the experimental mass differences ratio for the neutrino sector. Note that $\theta_{13} >0$ throughout the range.}
\label{lepang}
\end{figure}

Figure~\ref{lepang} shows all the angles obtained from the model consistent with the experimental ranges for angles and mass squared differences.

Since neutrinos in the model are Majorana, neutrino-less double beta decay ($0\nu\beta\beta$-decay) can take place. This decay is characterized by the (1,1) element of the neutrino mass matrix in the charged lepton mass basis which can be written as (see for example~\cite{BenTov:2011tj}).
\begin{equation}
m_{\beta\beta}=e^{2i\sigma}\cos^2\theta_{12}\cos^2\theta_{13}m_1+e^{2i\rho}\sin^2\theta_{12}\cos^2\theta_{13}m_2+\sin^2\theta_{13} m_3 \ ,
\label{mbb}
\end{equation}
where $m_j$ is the (real) mass of the $j$-th neutrino and $\sigma,\rho$ are Majorana phases. These masses can be parametrized using the mass squared differences and $m_3$ as
\begin{equation}
m_1=\sqrt{m_3^2+|\Delta m_{32}^2|-\Delta m_{21}^2}\text{, }\:\:\:m_2=\sqrt{m_3^2+|\Delta m_{32}^2|}.
\end{equation}

The present framework does not determine the absolute scale of neutrino masses and it is not possible to make a prediction for $m_{\beta\beta}$. Instead we only analyze the type of contribution that our model gives under the assumptions stated above that render the matrix elements in Eq.~(\ref{mnu}) of O(eV).

The current direct measurement upper bounds on $|m_{\beta\beta}|$ are given by~\cite{Bilenky:2011tr}
\begin{eqnarray}
|m_{\beta\beta}|&<&(0.20-0.32)\text{eV}\:\:\: (^{76}\text{Ge})\nonumber,\\
&<&(0.30-0.71)\text{eV}\:\:\: (^{130}\text{Te})\nonumber,\\
&<&(0.50-0.96)\text{eV}\:\:\: (^{130}\text{Mo}),
\end{eqnarray}
and future experiments expect to reach the $10^{-2}$~eV scale~\cite{Bilenky:2011tr}.

\begin{figure}[ht]
\includegraphics[width=0.40\textwidth]{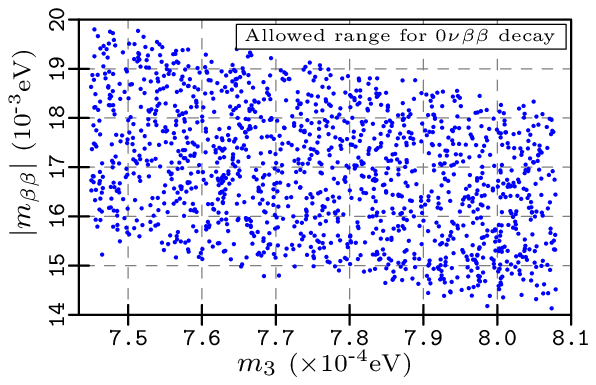} 
\includegraphics[width=0.45\textwidth]{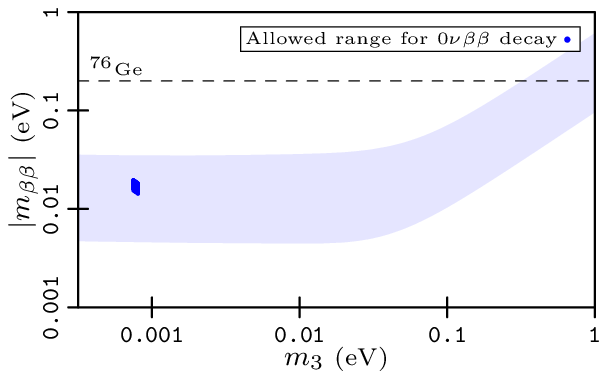}
\caption{{\it left}: Allowed range for the $0\nu\beta\beta$-decay in the $Q_4$ model. {\it right}: The light blue band corresponds to the allowed range for $|m_{\beta\beta}|$ from oscillation data at 3$\sigma$~\cite{Maltoni:2004ei}, the dashed line represents the lower limit from direct observation in $^{76}$\text{Ge} ($|m_{\beta\beta}|<0.20$~\cite{Bilenky:2011tr}), and the small dark blue spot is the allowed range in the left figure obtained from our model .}
\label{fig:vlessbdec}
\end{figure}

Using the angles in Figure~\ref{lepang}, the results are represented in the $|m_{\beta\beta}|$ - $m_{3}$ plane in the left plot of Figure~\ref{fig:vlessbdec}. We note that the texture in our model leads to the values $0$ and $\pi/2$ for $\sigma$ and $\rho$ respectively.

Upper and lower bounds can also be established from neutrino oscillation data~\cite{Hirsch:2006tt,Maltoni:2004ei}. The $3 \ \sigma$ allowed region for an inverted hierarchy is displayed in the light blue band on the right plot of Figure~\ref{fig:vlessbdec}, where we also show the small (dark blue spot) region corresponding to the present model.

%%%%%%%%%%%%%%%%%%%%%%%%%%%%%%%%%%%%%%%%%%%%%%%%%%%%%%%%%%%%%%%%%%%%%%%%%%%%%%%%%%%%%%%%
%%%%%%%%%%%%%%%%%%%%%%%%%%%%%%%%%%%%%%%%%%%%%%%%%%%%%%%%%%%%%%%%%%%%%%%%%%%%%%%%%%%%%%%%
%%%%%%%%%%%%%%%%%%%%%%%%%%%%%%%%%%%%%%%%%%%%%%%%%%%%%%%%%%%%%%%%%%%%%%%%%%%%%%%%%%%%%%%%

\section{Scalar potential}
\label{potential}
In this section we present the scalar potential and show that the vacuum alignment needed to generate the textures in Eq.~(\ref{QMM}) and Eq.~(\ref{mlfinal}) is consistent with its stability.

The gauge and flavor invariant potential is given by
\begin{equation}
V=V(H_{D})+V(H_{3})+V(H_{4})+V(\eta_{D})+V_{int}(H_{3},H_{4},H_{D},\eta_{D}) ,
\end{equation}
where
\begin{eqnarray}
V(H_{D})&=&\mu^{2}_{D}H_{D}^{\dagger}H_{D} + \ell_{1} \bigg\{ H_{D}^{\dagger}H_{D} \bigg\}^2 + \ell_{2}[\tilde{H}_{D}^{\dagger}H_{D}]_{\buno^{++}}[H_{D}^{\dagger}\tilde{H}_{D}]_{\buno^{++}} \\
\newline
V(H_{3})&=&\mu^{2}_{3}H_{3}^{\dagger}H_{3}+\ell_{3}(H_{3}^{\dagger}H_{3})^{2} \\
V(H_{4})&=&\mu^{2}_{4}H_{4}^{\dagger}H_{4}+\ell_{4}(H_{4}^{\dagger}H_{4})^{2} \\
V(\eta_{D})&=&\mu^{2}_{\eta}\eta_{D}^{\ast}\eta_{D} + \ell_{5} \bigg\{\eta_{D}^{\ast}\eta_{D}\bigg\}^2
+\ell_{6}\bigg\{\eta_{D}\eta_{D}\bigg\}\bigg\{\eta_{D}^{\ast}\eta_{D}^{\ast}\bigg\}
\end{eqnarray}
and where $V_{int}(H_{3},H_{4},H_{D},\eta_{D})$ is given by the sum of the following terms:
\begin{eqnarray}
V(H_{3},H_{4})&=&\ell_{7}|H_{3}^{\dagger}H_{4}|^2 + \ell_{8}(H_{3}^{\dagger}H_{3})(H_{4}^{\dagger}H_{4}) +\ell_{9}\left((H_{3}^{\dagger}H_{4})^2+h.c.\right) \nonumber \\
&+& \ell_{10}|\tilde{H}_{3}^{\dagger}H_{4}|^2 + \ell_{11}(\tilde{H}_{3}^{\dagger}H_{4})(H_{3}^{\dagger}\tilde{H}_{4})\\
\newline
V(H_{D},H_{3})&=&\ell_{12}|H_{D}^{\dagger}H_{3}|^2 + \ell_{13}[H_{D}^{\dagger}H_{D}]_{\buno^{++}}(H_{3}^{\dagger}H_{3}) + \ell_{14}|\tilde{H}_{D}^{\dagger}H_{3}|^2 \nonumber \\
&+& \ell_{15}(\tilde{H}_{D}^{\dagger}H_{3})(H_{D}^{\dagger}\tilde{H}_{3}) \\
\newline
V(H_{D},H_{4})&=&\ell_{16}|H_{D}^{\dagger}H_{4}|^2 + \ell_{17}[H_{D}^{\dagger}H_{D}]_{\buno^{++}}(H_{4}^{\dagger}H_{4}) + \ell_{18}|\tilde{H}_{D}^{\dagger}H_{4}|^2 \nonumber \\
&+& \ell_{19}(\tilde{H}_{D}^{\dagger}H_{4})(H_{D}^{\dagger}\tilde{H}_{4})
\end{eqnarray}
\begin{eqnarray}
V(H_{D},H_{3},H_{4})&=&\left( \ell_{20}(H_{D}^{\dagger}H_{3})(H_{4}^{\dagger}H_{D}) + \ell_{21}[H_{D}^{\dagger}H_{D}]_{\buno^{--}}(H_{3}^{\dagger}H_{4}) \right. \nonumber \\
&+& \left. \ell_{22}(H_{D}^{\dagger}H_{3})(H_{D}^{\dagger}H_{4}) \ell_{23}(\tilde{H}_{D}^{\dagger}H_{3})(H_{4}^{\dagger}\tilde{H}_{D}) \right. \nonumber \\
&+& \left. \ell_{24}(\tilde{H}_{D}^{\dagger}H_{3})(H_{D}^{\dagger}\tilde{H}_{4}) + \ell_{25}(H_{D}^{\dagger}\tilde{H}_{3})(\tilde{H}_{D}^{\dagger}H_{4}) + h.c. \right) \ \ \ \ 
\end{eqnarray}
\begin{eqnarray}
V(H_{D},\eta_{D})&=& \ell_{26} \bigg\{H_{D}^{\dagger}\eta_{D}\bigg\}\bigg\{\eta_{D}^{\ast}H_{D} \bigg\}
+ \ell_{27} \bigg\{ H_{D}^{\dagger}H_{D}\bigg\}\bigg\{\eta_{D}^{\ast}\eta_{D} \bigg\} 
\nonumber \\
&+& \ell_{28} \bigg\{ \tilde{H}_{D}^{\dagger}\eta_{D}\bigg\}\bigg\{\eta_{D}^{\ast}\tilde{H}_{D} \bigg\} 
+ \ell_{29} \bigg\{ H_{D}^{\dagger}\eta^{\ast}_{D}\bigg\}\bigg\{\eta_{D}H_{D} \bigg\} 
\end{eqnarray}
\begin{eqnarray}
V(H_{3},\eta_{D})&=& \ell_{30}(\eta_{D}^{\ast}\eta_{D})(H_{3}^{\dagger}H_{3})
+ \ell_{31} |H_{3}^{\dagger}\eta_{D}|^2 \\
\newline
V(H_{4},\eta_{D})&=& \ell_{32} (\eta_{D}^{\ast}\eta_{D})(H_{4}^{\dagger}H_{4})
+ \ell_{33} |H_{4}^{\dagger}\eta_{D}|^2 \\
\newline
V(H_{3}H_{4}\eta_D)&=&\ell_{34}(H_{3}^{\dagger}\eta_{D})(\eta_{D}^{\ast}H_{4})
+ \ell_{35}(\eta_{D}^{\ast}\eta_{D})(H_{3}^{\dagger}H_{4}) + h.c. \\
\newline
V(HH\eta)&=& \lambda_{1} H_{D}H_{3}\eta_{D} + \lambda_{2} H_{D}H_{4}\eta_{D} + h.c.
\end{eqnarray}

The terms inside the curly brackets correspond to the product of two $\bdos$'s (and so they contain four different $\buno$'s) and we include all possible combinations that - after multiplication of the two curly brackets - yield $\buno^{++}$.  Note that the last term $V(HH\eta)$ is the one in Eq.~(\ref{vHHeta}) where we included the SU(2) indexes explicitly. 

The SU(2) doublet scalar fields $H_i$ ($i=1,2,3,4$) are expressed as
\begin{eqnarray}
H_i = \left( \begin{array}{c}
H_i^+ \\
\frac{1}{\sqrt{2}}(v_i + h_i + i a_i)
\end{array} \right) \  \rightarrow
\langle H_i \rangle = \left( \begin{array}{c}
0 \\
\frac{v_i}{\sqrt{2}}
\end{array} \right) \ ,
\end{eqnarray}
where we work under the assumption that the vevs $v_i$ are real and thus the potential 
is CP-conserving.

The minimization of the potential gives the following relations:
\begin{eqnarray}
\mu_{D}^2&=&\frac{1}{2}\left(-4 \ell_{1}(v_{1}^{2} +v_{2}^{2}) - (\ell_{12} + \ell_{13}) v_{3}^{2} + 2\ell_{22}v_{3} v_{4} -(\ell_{16}+\ell_{17}) v_{4}^{2}\right), \\
\mu_{3}^{2}&=&\frac{1}{2 v_{3}}\left(-(\ell_{12} + \ell_{13}) (v_{1}^{2}+v_{2}^{2}) v_{3} - 2 \ell_{3} v_{3}^{3} + \ell_{22} (v_{1}^{2}+v_{2}^{2}) v_{4} - (\ell_{7} + \ell_{8} + 2 \ell_{9}) v_{3} v_{4}^{2}\right), \\
\mu_{4}^{2}&=&\frac{1}{2 v_{4}}\left(\ell_{22} (v_{1}^{2}+v_{2}^{2}) v_{3} -((\ell_{16} + \ell_{17})
(v_{1}^{2}+v_{2}^{2}) + (\ell_{7} + \ell_{8} + 2 \ell_{9}) v_{3}^{2}) v_{4} + 2 \ell_{4} v_{4}^{2}\right) ,
\end{eqnarray}
together with four massive scalar fields, three massive pseudoscalar fields, five massive charged scalar fields, and three massless Goldstone bosons. 

The vacuum alignment we need is $v_2 = 0 $ and all other vevs non-zero. Furthermore, the vevs must also satisfy the relation $v_1^2+v_2^2+v_3^2+v_4^2 = (246 \ \rm{GeV})^2$. Taking this into consideration we find that there are regions of parameter space where a stable minimum exist with masses in phenomenological acceptable ranges. A complete analysis of the scalar potential and its phenomenology is beyond the purpose of this paper and will be presented in a future publication.

%%%%%%%%%%%%%%%%%%%%%%%%%%%%%%%%%%%%%%%%%%%%%%%%%%%%%%%%%%%%%%%%%%%%%%%%%%%%%%%%%%%%%%%%
%%%%%%%%%%%%%%%%%%%%%%%%%%%%%%%%%%%%%%%%%%%%%%%%%%%%%%%%%%%%%%%%%%%%%%%%%%%%%%%%%%%%%%%%
%%%%%%%%%%%%%%%%%%%%%%%%%%%%%%%%%%%%%%%%%%%%%%%%%%%%%%%%%%%%%%%%%%%%%%%%%%%%%%%%%%%%%%%%

\section{Conclusions}

Renormalizable models of flavor provide an interesting alternative for the study of fermion masses and mixing. Furthermore, their scalar sector might involve interesting collider phenomenology due to the fact that the SU(2) {\it Higgs} doublets transform non-trivially under the flavor symmetry. 

Most constructions however require the introduction of a large number of SU(2) scalar doublets that make the phenomenological study cumbersome, except perhaps under some strong assumptions such as small interaction among the scalars and/or approximate diagonalizations and/or additional discrete Abelian symmetries. The purpose of this paper is to investigate if, and under what conditions, one can generate a renormalizable model with {\it just a few} SU(2) doublets.

The analysis is based on obtaining Fritzsch-like textures for the quark mass matrices. Once this is accomplished, the charged lepton mass matrix is in principle obtained by mimicking the down-type quarks. However, by fixing the transformation properties of the left-handed charged leptons, the left-handed neutrinos also get fixed. It turns out that in general it is not possible to obtain acceptable results for the neutrino sector, and one must consider alternative representations for the charged leptons that do not require the introduction of additional SU(2) doublets. We note that right-handed neutrinos are not present in the models and neutrino masses get generated radiatively.

We find that it is possible to construct renormalizable models of flavor with only four SU(2) doublet scalar fields transforming non-trivially under the flavor symmetry. The smallest group we found to work is $Q_4$. This is accomplished provided the following assumptions are met: i) there is an alignment between first and second generation Yukawa couplings (this is our strongest assumption), ii) there are no right-handed neutrinos and neutrino masses are generated radiatively, which requires the introduction of two SU(2) singlet scalar fields charged under both hypercharge and Lepton number, and iii) a particular (vacuum stable) vacuum alignment for the scalar sector must be imposed. We present a specific realization of such a model including the analysis for the vacuum stability of the scalar potential. The scalar phenomenology of the model is under investigation.

%%%%%%%%%%%%%%%%%%%%%%%%%%%%%%%%%%%%%%%%%%%%%%%%%%%%%%%%%%%%%%%%%%%%%%%%%%%%%%%%%%%%%%%%
%%%%%%%%%%%%%%%%%%%%%%%%%%%%%%%%%%%%%%%%%%%%%%%%%%%%%%%%%%%%%%%%%%%%%%%%%%%%%%%%%%%%%%%%
%%%%%%%%%%%%%%%%%%%%%%%%%%%%%%%%%%%%%%%%%%%%%%%%%%%%%%%%%%%%%%%%%%%%%%%%%%%%%%%%%%%%%%%%
\begin{acknowledgments}
A.~A. thanks The Abdus Salam ICTP for its hospitality while part of this work was carried out. This work was supported in part by CONACYT and SNI (Mexico).
\end{acknowledgments}

\appendix

%%%%%%%%%%%%%%%%%%%%%%%%%%%%%%%%%%%%%%%%%%%%%%%%%%%%%%%%%%%%%%%%%%%%%%%%%%%%%%%%%%%%%%%%
%%%%%%%%%%%%%%%%%%%%%%%%%%%%%%%%%%%%%%%%%%%%%%%%%%%%%%%%%%%%%%%%%%%%%%%%%%%%%%%%%%%%%%%%
%%%%%%%%%%%%%%%%%%%%%%%%%%%%%%%%%%%%%%%%%%%%%%%%%%%%%%%%%%%%%%%%%%%%%%%%%%%%%%%%%%%%%%%%

\section{Useful facts about $Q_4$}
\label{q4details}
The quaternion group $Q$, sometimes also called $Q_{4}$ or $Q_{8}$, has $8$ elements and five irreducible representations (irreps): $\buno^{++}$, $\buno^{+-}$, $\buno^{-+}$, $\buno^{--}$, and $\bdos$ (following notation in~\cite{Frigerio:2004jg,Frig1}) where the two-dimensional irrep is complex.

Let $A$ and $B$ be two two-dimensional irreducible representations of $Q_{4}$ such that $A=(\alpha_{1},\alpha_{2})$ and $B=(\beta_{1},\beta_{2})$. The following relations have been used in the paper (see~\cite{Frigerio:2004jg,Frig1,Q4rep}):
\begin{eqnarray}
A^*=(\alpha_{2}^{\ast},-\alpha^{\ast}_{1}) \ , 
\end{eqnarray}
\begin{eqnarray}
 \begin{array}{cc}
 \buno^{++} \otimes A=(\alpha_{1}, \alpha_{2}), & \buno^{+-}\otimes A = (\alpha_{2}, \alpha_{1}), \\
 \buno^{-+}\otimes A =(\alpha_{1}, -\alpha_{2}), & \buno^{--}\otimes A = (\alpha_{2},-\alpha_{1}) \ ,
 \end{array}
\end{eqnarray}
and
\begin{eqnarray}
A\otimes B= \buno^{++}\oplus \buno^{+-}\oplus \buno^{-+}\oplus\buno^{--},
\end{eqnarray}
where
\begin{eqnarray}
 \begin{array}{cc}
 \buno^{++}\sim (\alpha_{1}\beta_{2}-\alpha_{2}\beta_{1}), & \buno^{+-}\sim (\alpha_{1}\beta_{1}-\alpha_{2}\beta_{2}), \\
 \buno^{-+}\sim (\alpha_{1}\beta_{2}+\alpha_{2}\beta_{1}), & \buno^{--}\sim (\alpha_{1}\beta_{1}+\alpha_{2}\beta_{2}). \
 \end{array}
\end{eqnarray}

%%%%%%%%%%%%%%%%%%%%%%%%%%%%%%%%%%%%%%%%%%%%%%%%%%%%%%%%%%%%%%%%%%%%%%%%%%%%%%%%%%%%%%%%
%%%%%%%%%%%%%%%%%%%%%%%%%%%%%%%%%%%%%%%%%%%%%%%%%%%%%%%%%%%%%%%%%%%%%%%%%%%%%%%%%%%%%%%%
%%%%%%%%%%%%%%%%%%%%%%%%%%%%%%%%%%%%%%%%%%%%%%%%%%%%%%%%%%%%%%%%%%%%%%%%%%%%%%%%%%%%%%%%

\section{Yukawa textures}
\label{textures}

The analysis in this paper is based on obtaining the Fritzsch-like textures for the quark mass matrices $M_{u,d}$
\begin{eqnarray}\label{fritzsch}
M_{u,d} = \left( \begin{array}{ccc}
0 & A_{u,d} & 0 \\
A_{u,d} & 0 & B_{u,d} \\
0 & D_{u,d} & C_{u,d}
\end{array} \right) \ .
\end{eqnarray}

However, there are related textures that can also be used in our scenario. To see this we write the CKM-matrix as $V_{CKM} = V_{Lu} V_{Ld}^{\dagger}$, where $V_{L(u,d)}$ are the unitary matrices that diagonalize the {\it squared} quark mass matrices
\begin{eqnarray}
V_{L(u,d)}M_{u,d}M^{\dagger}_{u,d}V^{\dagger}_{L(u,d)}=\operatorname{diag}(m^{2}_{u,d},m^{2}_{c,s},m^{2}_{t,b}).
\end{eqnarray}

Denoting by $M_{u,d}^F \equiv M_{u,d}M_{u,d}^{\dagger}$ and using Eq.~(\ref{fritzsch}) we see that
\begin{eqnarray}\label{m2f}
M_{u,d}^F = \left( \begin{array}{ccc}
A_{u,d}^{2}         &            0            & A_{u,d}D_{u,d} \\
0                   & A_{u,d}^{2}+B_{u,d}^{2} & B_{u,d}C_{u,d} \\
A_{u,d}D_{u,d}      & B_{u,d}C_{u,d}          & C_{u,d}^{2}+D_{u,d}^{2}
\end{array} \right) \ .
\end{eqnarray}

The relevant observation is that any matrices $M_u$ and $M_d$ whose {\it squares} give the matrices $M^F_u$ and $M^F_d$, respectively, will then lead to the same CKM matrix (up to the phases introduced in Eq.~(\ref{quarkphases})). The following matrices have this property
\begin{eqnarray}\label{matrices}
\left( \begin{array}{ccc}
0 & 0 & A \\
A & B & 0 \\
0 & C & D
\end{array} \right), \ \
\left( \begin{array}{ccc}
A& 0 & 0 \\
0 & A & B \\
D & 0 & C
\end{array} \right),  \
\left( \begin{array}{ccc}
A & 0 & 0 \\
0     & B & A \\
D & C &  0
\end{array} \right), \
\left( \begin{array}{ccc}
0  & 0 & A \\
B & A & 0 \\
C & 0 & D
\end{array} \right),  \ 
\left( \begin{array}{ccc}
0 & A & 0 \\
B & 0 & A \\
C & D & 0
\end{array} \right),  \ 
\end{eqnarray}
and so it is conceivable that models with $Q_4$ - and other symmetries - can be constructed that lead to some of these quark mass matrices. For example, if the first and second generation right-handed up-type quark assignments for the model presented in the paper were reversed and put in the {\it normal} order (see Eq.~(\ref{transformations})), i.e. $(u_{R1} \  u_{R2})$, then the mass matrix $M_u$ would take the form of the first matrix in Eq.~(\ref{matrices}). Note that the matrices in Eq.~(\ref{matrices}) correspond to all possible column interchanges of the matrix in Eq.~(\ref{fritzsch}). Regarding the phase factorization Eq.~(\ref{quarkphases}) we find that only the last matrix above gets factorized in exactly the same way as the Fritzcsh-like textures, while the rest require additional assumptions such as $A \in \mathbb{R}$ and/or $B \in \mathbb{R}$. We stress that the minimum number of SU(2) doublets needed to generate any of these matrices (in both the up and down-type quark sectors) is four.

It is important to note that the matrices above do not represent the only possibility for generating an acceptable CKM matrix. They are simply variations of the Fritzsch-like matrix Eq.~(\ref{fritzsch}) that satisfy Eq.~(\ref{m2f}). Interesting alternatives do exist. See for example the recent work in~\cite{texs} where it is shown that having $M_u$ similar to Eq.(~\ref{fritzsch}) and an $M_d$ given by
\begin{eqnarray}
M_{d} = \left( \begin{array}{ccc}
0     &   A_{d} & 0 \\
A^{\ast}_{d}  & B_{d} & 0 \\
0     &     0   & C_{d}
\end{array} \right),
\end{eqnarray}
leads to an acceptable CKM matrix. This type of texture, although not of the form in Eq.~(\ref{m2f}), can also be obtained from $Q_4$ with a minimum of four SU(2) doublets.

%%%%%%%%%%%%%%%%%%%%%%%%%%%%%%%%%%%%%%%%%%%%%%%%%%%%%%%%%%%%%%%%%%%%%%%%%%%%%%%%%%%%%%%%
%%%%%%%%%%%%%%%%%%%%%%%%%%%%%%%%%%%%%%%%%%%%%%%%%%%%%%%%%%%%%%%%%%%%%%%%%%%%%%%%%%%%%%%%
%%%%%%%%%%%%%%%%%%%%%%%%%%%%%%%%%%%%%%%%%%%%%%%%%%%%%%%%%%%%%%%%%%%%%%%%%%%%%%%%%%%%%%%%

\section{Radiative neutrino masses}\label{neutrinos}
In absence of right-handed neutrinos the only possible mass terms for left-handed neutrinos are Majorana mass terms. The simplest mass term in this case, without the introduction of scalars with non trivial SU(2) representations, is
the dimension five operator with form $\mathcal{L} \propto \overline{l}_L^c l_L \frac{HH}{M}$. Although this term is non-renormalizable, it may be induced by radiative corrections if we introduce a scalar field that breaks lepton 
number (provided there are at least two SU(2) Higgs doublets \cite{Zee}). This is why we have introduced the fields $\eta_1$ and $\eta_2$ in our renormalizable model. 

In order to see how this works, consider the following example: A two Higgs doublet model with SM fermion content and an additional scalar field $h$ with charges $(1,-1)$ under $SU(2)\times U(1)_Y$ and lepton number $L=2$~\cite{Zee,Fukugita:2003en}. The Yukawa couplings of $h$ are
\begin{eqnarray}\label{Lyuk}
    \mathcal{L}_{llh}=\kappa^{ab}\epsilon_{ij}\overline{(L^a_{i})^c}L^b_{j}h^{*}+
    h.c. \ ,
\end{eqnarray}
where $i,j$ are SU(2) indices, $a,b$ are family indices, $\kappa^{ab}=-\kappa^{ba}$ from Fermi statistics, and $L_i$ denotes the SU(2) lepton doublets. 
If there are two (or more) Higgs doublets, there will be a cubic coupling term like
\begin{eqnarray}\label{Lcubic}
    \mathcal{L}_{HHh}=\lambda_{\alpha\beta}\epsilon_{ij}H_i^{\alpha}H_j^{\beta}h + h.c.,
\end{eqnarray}
with $\lambda_{\alpha\beta}=-\lambda_{\beta\alpha}$, and $\alpha,\beta=1,2$. This term explicitly violates
lepton number and allows the generation of majorana masses for the neutrinos.

\begin{figure}[ht]
\includegraphics[width=10cm]{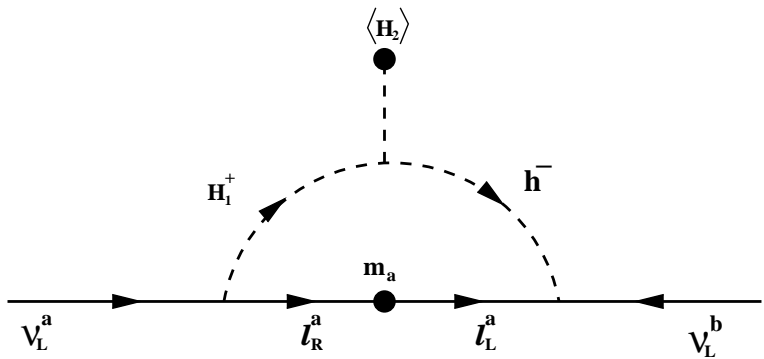}
\caption{One loop diagram giving rise to neutrino Majorana mass.}
\label{fig:loop}
\end{figure}

Notice that Eqs.~(\ref{Lyuk}) and (\ref{Lcubic}) (together with the usual Yukawa term from the lepton sector)
lead to the diagram shown in figure~\ref{fig:loop} that contributes
to a Majorana mass term as
\begin{eqnarray}\label{Mab}
M_{ab}=(-1)\kappa^{ab}m_a^2\frac{\lambda_{12}v_2}{v_1}\frac{1}{(4\pi)^2}\frac{1}{m_{H_1}^2-m_h^2
    }\log\frac{m_{H_1}^2}{m_h^2},
    \end{eqnarray}
where $m_{H_1}$ denotes the charged Higgs mass and $m_h$ the mass of the singlet 
field $h$.\footnote{We note that this is not yet in the scalar mass basis since the $H^0 H^{+}h^{-}$ term induces mixing between $H^{+}$ and $h^{-}$ However, we expect $m_H << m_h$, and work in the approximation that treats $m_h$ and $m_{H_1}$ as the physical masses.} Thus, the total contribution, including the diagram with $\nu^b_L$ and $\nu^a_L$ interchanged 
(which has the same form as in Eq.~(\ref{Mab}) but with $a\leftrightarrow b$) is
\begin{eqnarray} \label{mab}\nonumber
m_{ab} &=& \kappa^{ab}(m_b^2-m_a^2)\frac{\lambda_{12}v_2}{v_1}\frac{1}{(4\pi)^2}\frac{1}{m_{H_1}^2-m_h^2
    }\log\frac{m_{H_1}^2}{m_h^2} \\
       &=& \kappa^{ab}(m_b^2-m_a^2)\frac{\lambda_{12}
    v_2}{v_1}F(m_H^2,m_{h}^2),	
    \end{eqnarray}
with~\cite{Fukugita:2003en}
\begin{eqnarray}\label{F}
    F(x,y)=\frac{1}{16\pi^2}\frac{1}{x-y}\log\frac{x}{y} \ .
\end{eqnarray}

Note that in this example the antisymmetry of $\kappa$ forbids diagonal mass matrix elements. The non-trivial representations of neutrinos and scalars under the flavor symmetry can alter this situation.

%%%%%%%%%%%%%%%%%%%%%%%%%%%%%%%%%%%%%%%%%%%%%%%%%%%%%%%%%%%%%%%%%%%%%%%%%%%%%%%%%%%%%%%%
%%%%%%%%%%%%%%%%%%%%%%%%%%%%%%%%%%%%%%%%%%%%%%%%%%%%%%%%%%%%%%%%%%%%%%%%%%%%%%%%%%%%%%%%
%%%%%%%%%%%%%%%%%%%%%%%%%%%%%%%%%%%%%%%%%%%%%%%%%%%%%%%%%%%%%%%%%%%%%%%%%%%%%%%%%%%%%%%%

\end{document}